\documentclass[sigconf]{acmart}

\AtBeginDocument{%
  }

\setcopyright{acmlicensed}
\copyrightyear{2025}
\acmYear{2025}
\acmDOI{XXXXXXX.XXXXXXX}

\acmConference[]{}{}

\usepackage{booktabs}
\usepackage{tabularx}
\usepackage{makecell}
\usepackage{multirow}
\usepackage{enumitem}

\begin{document}

\title[Imposing View Time Limits in Crowdsourced Image Classification]{Towards Fair Pay and Equal Work: \\ Imposing View Time Limits in Crowdsourced Image Classification}

\author{Gordon Lim}
\email{gbtc@umich.edu}
\affiliation{%
  \institution{University of Michigan}
  \city{Ann Arbor}
  \state{Michigan}
  \country{USA}
}

\author{Stefan Larson}
\email{stefan.larson@vanderbilt.edu}
\affiliation{%
  \institution{Vanderbilt University}
  \city{Nashville}
  \state{Tennessee}
  \country{USA}}

\author{Yu Huang}
\email{yu.huang@vanderbilt.edu}
\affiliation{%
  \institution{Vanderbilt University}
  \city{Nashville}
  \state{Tennessee}
  \country{USA}}
  
\author{Kevin Leach}
\email{kevin.leach@vanderbilt.edu}
\affiliation{%
  \institution{Vanderbilt University}
  \city{Nashville}
  \state{Tennessee}
  \country{USA}}
  
\renewcommand{\shortauthors}{Lim et al.}

\begin{abstract}
    Crowdsourcing is a common approach to rapidly annotate large volumes of data in machine learning applications. 
    Typically, crowd workers are compensated with a flat rate based on an estimated completion time to meet a target hourly wage.
    Unfortunately, prior work has shown that variability in completion times among crowd workers led to overpayment by 168\% in one case, and underpayment by 16\% in another.
    However, by setting a time limit for task completion, it is possible to manage the risk of overpaying or underpaying while still facilitating flat rate payments.
    In this paper, we present an analysis of the impact of a time limit on crowd worker performance and satisfaction. 
    We conducted a human study with a maximum view time for a crowdsourced image classification task.
    We find that the impact on overall crowd worker performance diminishes as view time increases.
    Despite some images being challenging under time limits, a consensus algorithm remains effective at preserving data quality and filters images needing more time. 
    Additionally, crowd workers' consistent performance throughout the time-limited task indicates sustained effort, and their psychometric questionnaire scores show they prefer shorter limits.
    Based on our findings, we recommend implementing task time limits as a practical approach to making compensation more equitable and predictable.
    Our code and data are available at \url{https://github.com/gordon-lim/sdogs-10h}.
\end{abstract}



\received{14 August 2024}

\maketitle

\section{Introduction}
Supervised deep learning methodologies demand large amounts of annotated data~\cite{deep-net, vit}.
Crowdsourcing has emerged as a popular method of data annotation due to its cost-efficiency~\cite{mathew2021hatexplain, crwdsrc-survey, imagenet}. 
The flat rate compensation model is widely used for its convenience, being the default option on platforms like Amazon Mechanical Turk, and due to the tradition of compensating research participation based on output \cite{the-problem-of-flat-rates}.
However, as crowdsourcing can be a full time job for some workers, they need a minimum wage to afford a decent standard of living~\cite{the-problem-of-flat-rates, posch2018characterizing}.
As such, a recent movement towards giving crowd workers a fair pay has called on researchers to consider estimated hourly wages for completing their tasks~\cite{beyond-fair-pay, the-problem-of-flat-rates}.
For instance, venues like NeurIPS\footnote{https://nips.cc/public/EthicsGuideline} and ACL\footnote{https://aclrollingreview.org/responsibleNLPresearch} ask that researchers using crowdsourced labor disclose estimated hourly wages in their papers.  
A common way to estimate completion times is by conducting a pilot study and measuring an average completion time~\cite{daemo, measuring-crwdsrc-effort}.
However, the variability in individual completion times means that there are cases of overcompensation and undercompensation~\cite{the-problem-of-flat-rates, fair-work, crwdsrc-ethical-issues}.
For example, \citet{the-problem-of-flat-rates} show that in one study, which paid a flat rate to honor a local minimum wage using average estimated time during a pilot study, the requesters paid 168\% more than they would have if they had used the actual average time taken to complete the tasks. 
In another case, they paid 16\% less than the intended minimum hourly wage. 
Consequently, there is a need to develop more equitable compensation schemes.

\begin{figure}
    \centering
    \includegraphics[width=\linewidth]{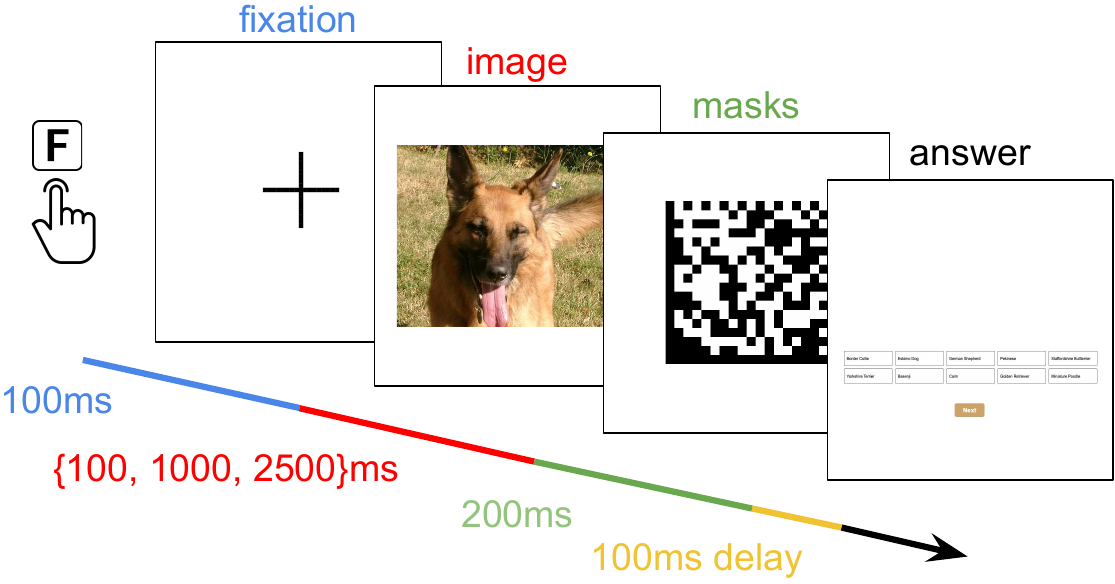}
    \caption{Experimental setup of view time limit.  
    In this paper, we investigate 100/1000/2500ms time limits to examine their impact on data quality and worker experience.}
    \label{fig:view-time-limit}
\end{figure}

\citet{the-problem-of-flat-rates} recommends that, instead of a flat rate, workers should be paid based on the time they spend on tasks.
\citet{fair-work} developed an algorithm that uses self-reported estimates from crowd workers to automatically grant post-hoc bonuses to those falling below minimum wage. 
However, these time-adjusted models implicitly trust crowd workers, who are incentivised to intentionally inflate their completion times. 
Consequently, these methods have not been widely adopted, and instead the flat rate compenstation scheme remains dominant due to its convenience and ease of budgeting.
The crowdsourcing platform Prolific\footnote{\url{https://prolific.com}} uses the median completion time of all submissions\footnote{\url{https://researcher-help.prolific.com/hc/en-gb/articles/360019777180-How-do-I-resolve-underpaying-studies}} to recommend fairer wages \cite{NEURIPS2022_1968ea7d}. 
However, even with median-based estimates, there remains the issue of overpayment and underpayment at the individual level~\cite{the-problem-of-flat-rates}.

In light of this discussion, we propose imposing time limits on crowdsourcing tasks for the following reasons:
(1) to help workers save time and manage expectations while ensuring they receive fair pay, 
(2) to assist requesters in developing a fair and consistent payment strategy that is easy to implement, and 
(3) to prevent overcompensation and save costs.
In this paper, we focus on image classification crowdsourcing tasks to compare our findings with previous work that did not use time limits~\cite{cifar10h}. 
One challenge with proposing a time limit is to address cognitive costs rather than psychomotor costs~\cite{measuring-crwdsrc-effort}, ensuring we do not discredit workers who have already begun cognitive processing (i.e., studying the image to identify it).
To address this, we set a time limit on how long crowd workers can view the image but do not restrict the time needed to physically submit their response (Figure~\ref{fig:view-time-limit}). 

For this approach to be widely adopted, it must also have the following two desirable properties: (1) it does not negatively impact data quality, and (2) it does not negatively impact worker sentiment, as a less desirable task might deter workers or lead to expectations of higher compensation. 
In this paper, we seek to validate these properties and guide the establishment of view time limits as a practical approach for addressing ethical concerns surrounding fair compensation and transparency over task completion times.

\begin{figure}[t]
    \centering
    \includegraphics[width=0.8\linewidth]{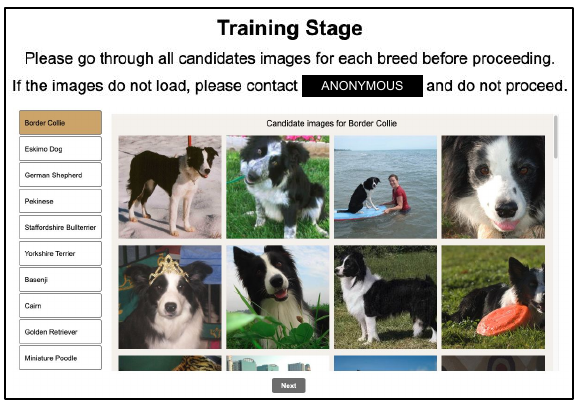}
    \caption{Training procedure. Participants are shown examples of each class.  They proceed after seeing all images in all classes.}
    \label{fig:train-screenshots}
\end{figure}

\begin{figure}[t]
    \centering
    \includegraphics[width=0.8\linewidth]{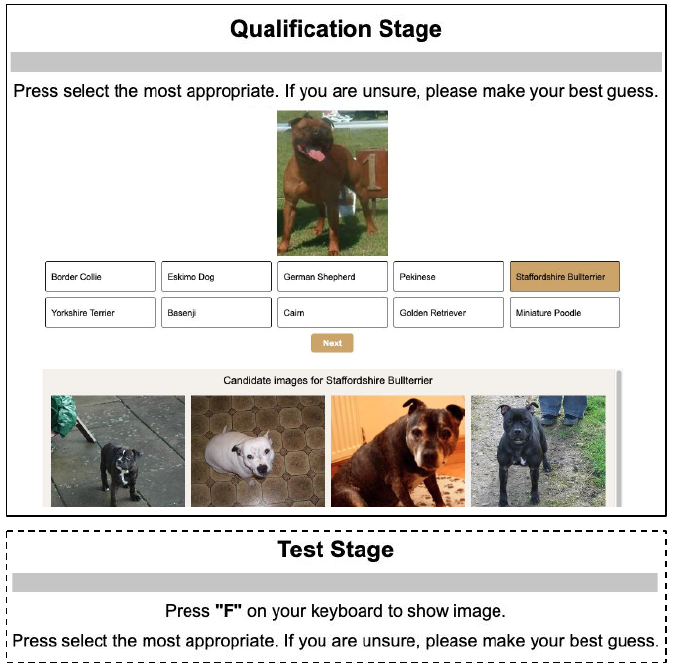}
    \caption{Qualification and Test trials.
    Participants are shown a dog image and must select the best breed category. 
    In the qualification stage, there is no time limit. 
    In the time-limited test stage, the image disappears after 100ms, 1000ms, or 2500ms.
    Participants can revisit training images by clicking each category. 
    A grey bar shows overall progress.}
    \label{fig:qual-test-screenshots}
\end{figure}

\section{Related Work}

\paragraph{Compensation Models}
Many crowdsourcing compensation models have been explored. 
\citet{the-problem-of-flat-rates} provides three categories of these approaches.
First, the  \emph{flat-rate model}, where all participants are paid the same for completing a task~\cite{liu2022audio, mallya2022implicit, gong2022explicable, lu2022learn}.
Second, the \emph{time-adjusted model}, where pay is based on the time spent~\cite{fair-work}.
Third, the  \emph{quality-adjusted model}, which tries to incentivize quality work~\cite{paying-collab-work, peer-truth, peer-dependent-reward}.
Despite the benefits of time- and quality-adjusted models, the flat-rate model remains the most popular due to its convenience as the default option on most crowdsourcing platforms and the ease of budgeting~\cite{the-problem-of-flat-rates}.
Yet, \citet{the-problem-of-flat-rates} show that a wide variability of completion times among crowd workers leads to instances of over- and undercompensation~\cite{fair-work, paying-collab-work}.
In this paper, we consider time limits imposed on crowdsourcing tasks to allow researchers to retain the convenience of flat rate compenstation schemes while controlling the variability of individual completion times. 
This essentially fixes the level of effort expected from crowd workers, therefore making fixed compensation more likely to fairly reflect their time commitment.

\paragraph{Time Limits and Cognition}

The impacts of time limits as a source of time pressure on cognition have been widely explored in the field of psychology~\cite{WU2018244, chajut2003selective, ferrari2001procrastination, walczyk1999time}. 
\citet{chajut2003selective} showed that selective attention, evaluated using the Stroop test, improved when the time limit was reduced. 
In another study on online search performance, \citet{WU2018244} found that the use of a time limit did not negatively affect overall performance but minimized distractive browsing with shorter time limits. 
On the other hand, \citet{walczyk1999time} observed that students' reading comprehension improved under mild time pressure but declined when the pressure was severe. 
Additionally, \citet{ferrari2001procrastination} discovered that chronic procrastinators tend to be slower and make more errors when subjected to time limits, suggesting that different personalities react to time limits differently. 
These findings collectively highlight the complex nature of time pressure on cognitive functions, where careful consideration is needed regarding the use of time limits.

\paragraph{Time Limits in Crowdsourcing}

Within crowdsourcing, time limits have been used by \citet{time-limit}, \citet{measuring-crwdsrc-effort} and \citet{maddalena2016crowdsourcing}.

\citet{time-limit} investigated different input elicitation methods to improve image classification by crowd workers, implementing a 60-second view time limit to prevent cognitive overload. 
Participants could submit their answers at any point during the 60-second period or afterward. 
Since their focus was on the elicitation methods, the impact of the 60-second limit on data quality was not explored.

\citet{measuring-crwdsrc-effort} tested various time limits to estimate hourly wages based on the minimum time needed for quality data. 
Crowd workers' submissions were disabled after the time limit.
Their investigation focused on the use of time limits in a pre-deployment assessment to estimate completion time, hence it was acceptable to treat incomplete submissions as incorrect.
In our study, limiting only image view time in classification tasks still allows submissions based on the cognitive processing completed.
As we consider the practicality of using time limits in data annotation processes, our study also analyzes the direct effects of time pressure on performance.

Most closely related to our study is the work by \citet{maddalena2016crowdsourcing}, who analyzed the effects of various time limits on crowdsourced relevance judgments in information retrieval tasks. 
In their setup, a target document would disappear after a set time limit. 
Due to the similarities between our work and theirs, we highlight three key differences in ours.
First, the fundamental difference in modality (images vs. text) suggests that different cognitive processes are engaged, which may lead to varying effects under time limits. Second, while \citet{maddalena2016crowdsourcing} focused on optimizing time limits to improve cost-efficiency for the requester, our work also considers the benefits for the crowd workers. To this end, we gather additional feedback from participants, including psychometric scores, to better understand their experience.
Third, although \citet{maddalena2016crowdsourcing} also employed a majority vote consensus, we extend the analysis by examining the potential impacts of combining a consensus approach with varying time limits. 
Furthermore, our study publicly releases our participant data to support research that requires datasets with a full human label distribution, such as learning with noisy labels \cite{plank-2022-problem, cifar10h}.

\section{Research Questions}

To adopt view time limits as a means to promote ethical and fair compensation and task transparency, it is crucial to first understand their impact on crowd worker performance and, consequently, on data quality.
Additionally, identifying the types of images that are particularly challenging under a time limit will help requesters recognize the limitations of such a constraint for their dataset. 
Furthermore, a view time limit should not hinder the effectiveness of majority vote consensus due to the influence of "fast and frugal" cognitive heuristics~\cite{homo-heuristicus} under a time limit.
Finally, a view time limit should not negatively affect how crowd workers perceive the task, as a less appealing task could discourage workers from working the task.
Based on these considerations, we pose the following five research questions:
\begin{enumerate}[label=\textbf{RQ\arabic*:}]
    \item How does time limit impact \emph{individual} participant accuracy?
    \item What is the trade-off between \emph{overall} performance accuracy and varying time limits?
    \item Which images are more difficult under a view time limit?
    \item How can consensus algorithms mitigate the impact of time limits on performance accuracy?
    \item How do view time limits impact crowd worker satisfaction and perceived effort during the task?
\end{enumerate}

\section{Methods}

In this section, we present a human study protocol we used to answer our aforementioned research questions. 
We recruited participants via Prolific to complete an image classification task with different time limits imposed.
Specifically, we discuss the Stanford Dogs dataset~\cite{stanford-dogs}, from which we built our task.
Then, we describe the recruitment process as well as the design of our human study survey instrument for collecting performance data about each participant.

\subsection{Dataset}

In our study, we consider a high-resolution image classification benchmark~\cite{wei2021fine, van2021benchmarking, yin2020fine}.
We selected the Stanford Dogs dataset~\cite{stanford-dogs}, which comprises 20,580 images spanning 120 distinct dog breeds, and is a subset of ImageNet~\cite{imagenet}.
Unlike ImageNet, which has been found to contain many label errors~\cite{cleanlab, northcutt2021labelerrors}, the Stanford Dogs images have been meticulously verified by~\citet{stanford-dogs} against images on Wikipedia and images within the same category.
To prevent overwhelming participants and minimize biases such as selection bias or fatigue, we deliberately chose a subset of 10 breed categories based on visual similarity following~\citet{dodge2017study}.
The chosen categories are \textit{Border Collie, Eskimo Dog, German Shepherd, Pekinese, Staffordshire Bullterrier, Yorkshire Terrier, Basenji, Cairn, Golden Retriever,} and \textit{Miniature Poodle}.
While~\citet{dodge2017study} selected the \textit{Dalmatian} category from ImageNet, we randomly chose \textit{Cairn} as a substitute because \textit{Dalmatian} is absent from the Stanford Dogs dataset.

In our time-limited test, participants were shown 25 randomly sampled images per breed and asked to identify the correct breed after viewing each image.
To prepare each participant, they were first trained with a random sample of 50 images per breed, which they could page through to see diverse representations without any time limit.
Then, participants completed a qualification task, correctly identifying 3 randomly sampled images per breed, also without a time limit, to ensure they could discriminate between visually similar breeds and filter out low-quality workers, following practices established in prior work~\cite{butala2024promise, hwang2023memecap, dodge2017study}.
In total, we used 500 training images, 30 qualification images, and 250 test images across 10 breed categories.

\renewcommand{\arraystretch}{1.5}
\renewcommand{\cellalign}{l}
\begin{table*}[t]
    \centering 
    \small 
    \caption{Overview of participants' survey responses.}
    \begin{tabularx}{\linewidth}{|p{2cm}|c|c|c|X|X|}
        \hline 
        \thead{\textbf{Topics}} & \multicolumn{3}{c|}{\thead{\textbf{Freq by Cohort (ms)}}} & \thead{\textbf{Description}} & \thead{\textbf{Examples}} \\
        \cline{2-4}
        & \textbf{100} & \textbf{1000} & \textbf{2500} & & \\
        \hline 
        Positive \newline comments & 3 & 2 & 2 & Participant left positive remarks on the study & ``Very easy and fun" \newline ``While I understood the limited time, I still enjoyed this task." \\ 
        \hline 
        Too slow & 0 & 0 & 4 & Participant felt the time limit was excessive and wanted to proceed faster & ``I think I had plenty of time, if anything less time and delay between choosing." \\
        \hline 
        Too quick & 6 & 4 & 3 & Participant commented on time limit being too short & ``It was a bit difficult to catch the photos in time." \newline ``time was too limited on some pictures" \\
        \hline 
        Cairn vs. \newline Yorkshire & 3 & 2 & 2 & Participant felt that Cairn and Yorkshire terrier were too visually close & ``cairns and yorkys are especially difficult to distinguish." \newline ``Some of the breeds looked similar, Cairn and Yorkies mostly" \\ 
        \hline
    \end{tabularx}
    \label{tab:code-book}
\end{table*}

\subsection{Recruitment}

We recruited 30 participants via the crowdsourcing platform Prolific~\cite{douglas2023data, Abbas_Gadiraju_2022}.
The study received Institutional Review Board exemption (ID: 231169).
Participation was restricted to individuals in the United States, and participants were required to use a desktop computer for the study and to complete the study in one sitting. 
To avoid biasing our results, we did not inform participants about the different cohorts. 
As such, we based our estimated hourly wage on the maximum view time limit of 2500ms. 
We conducted a pilot study with six volunteers recruited from research lab members and personal contacts before publishing the study on Prolific and estimated that the entire study, including reading our instructions, training, qualification, and the post-study survey (which have no time limits), would take under 45 minutes.
Therefore, we compensated participants \$12 for completing the study, ensuring at least a wage of \$15 per hour for all participants.

\subsection{Procedure}

Participants would select our study from the Prolific page, which led to a custom URL where we set up a Flask-based stimulus website containing our image classification task. 
Each participant is assigned a unique ID by Prolific upon entering the study, and we use the ID to manage each participant's answers. 
Upon enrolling, participants were randomly and evenly assigned into a time limit group --- 100ms, 1000ms, or 2500ms. 
These time limits were chosen based on our pilot studies to explore the impact of very short and reasonably long limits.

\subsubsection{Training and Qualifying Participants}
To help prepare participants to complete our time-limited test, we provide a training task.
In the training stage (Figure \ref{fig:train-screenshots}), participants review all training images and can only proceed after scrolling to the end and viewing all breeds that will appear later. 
Next, we use a qualification task to help filter out low-quality workers. 
In the qualification stage (Figure~\ref{fig:qual-test-screenshots}), participants must correctly identify 27 out of 30 randomly selected dog images to continue with the study. 
There is no time limit during this stage. 
Participants are informed that failure to meet the accuracy criterion will prompt them to ``return" their Prolific submission, following Prolific's protocol for incomplete studies.
The training and qualification stages take approximately 5 minutes to complete.
Upon successful qualification, participants will advance to the test-limited test.

\subsubsection{Time-Limited Test}
Figure~\ref{fig:view-time-limit} illustrates our setup for our view time limit, and Figure~\ref{fig:qual-test-screenshots} shows a screenshot of the classification task.
Participants were shown a single image at a time, following the view time limit process similar to~\citet{adversarial-time-limited}.
First, they press the `F' key to initiate showing a fixation cross for 100ms.
Second, an image of a dog appears for 100ms, 1000ms, or 2500ms, based on the participant's randomly-assigned cohort. 
After the given time had passed, the dog image would disappear.
To prevent further cognitive processing of the image, we showed a sequence of ten mask images, each lasting 20ms~\cite{adversarial-time-limited}.
Next, buttons for each dog breed would become clickable, allowing the participant to select the best option. 
We stored the participant's selection for each image.

\subsubsection{Qualitative Participant Data}

Following the test stage, participants are directed to a brief Qualtrics survey to collect demographic information and qualitative responses.
The survey includes open-ended questions on task difficulty, challenging image types, and general feedback.
The survey responses were collected, coded, and organized for analysis.
Participants also complete the Positive and Negative Affect Schedule (PANAS) questionnaire~\cite{panas}, a self-report measure of affect, rating how they generally feel on average for positive and negative emotion words like \textit{excited}, \textit{interested}, \textit{upset}, and \textit{irritable}, allowing us to measure their post-study sentiment.

\subsubsection{SDOGS-10H Dataset} 

Based on the steps described above, we release a dataset from our study (called \emph{SDOGS-10H}), consisting of 7500 human label annotations over 250 dog images sourced from Stanford Dogs. 
These annotations were gathered from participants who viewed the images for durations of 100ms, 1000ms, or 2500ms. 
We also include participants' responses from the post-study survey.
Next, we discuss the results of analyzing this data we collected.

\section{Results and Analysis}

Our human study protocol facilitated the collection of image classification labels by crowd workers. 
We investigate the impact of varying time limits on crowd worker performance and satisfaction. 
Our analyses include a comparison of our results with that of the CIFAR-10H study~\cite{cifar10h}, which involved a similar image classification task with no time limit.
CIFAR-10H comprises over 500,000 human-labeled images from the CIFAR-10 test set~\cite{cifar10}, collected from 2,571 participants recruited on Amazon Mechanical Turk. 
Unlike our study, CIFAR-10H used low-resolution images and broad categories (e.g., \textit{frog} and \textit{airplane}), potentially limiting its real-world relevance. 
Additionally, our study differentiates between decision-making and response input time, by separately recording time for viewing the image and physically submitting their answer. 
To address this disparity, we make a simplifying assumption that decision-making and response input occurred simultaneously in the CIFAR-10H study. 
Consequently, we treat the time data in CIFAR-10H and the duration of our view time limit equally.
As part of our data cleaning process, we removed any erroneous entries from CIFAR-10H that recorded negative times.

\paragraph{RQ1: How does time limit impact individual participant accuracy?}

\begin{figure}[t]
    \centering
    \includegraphics[width=\linewidth]{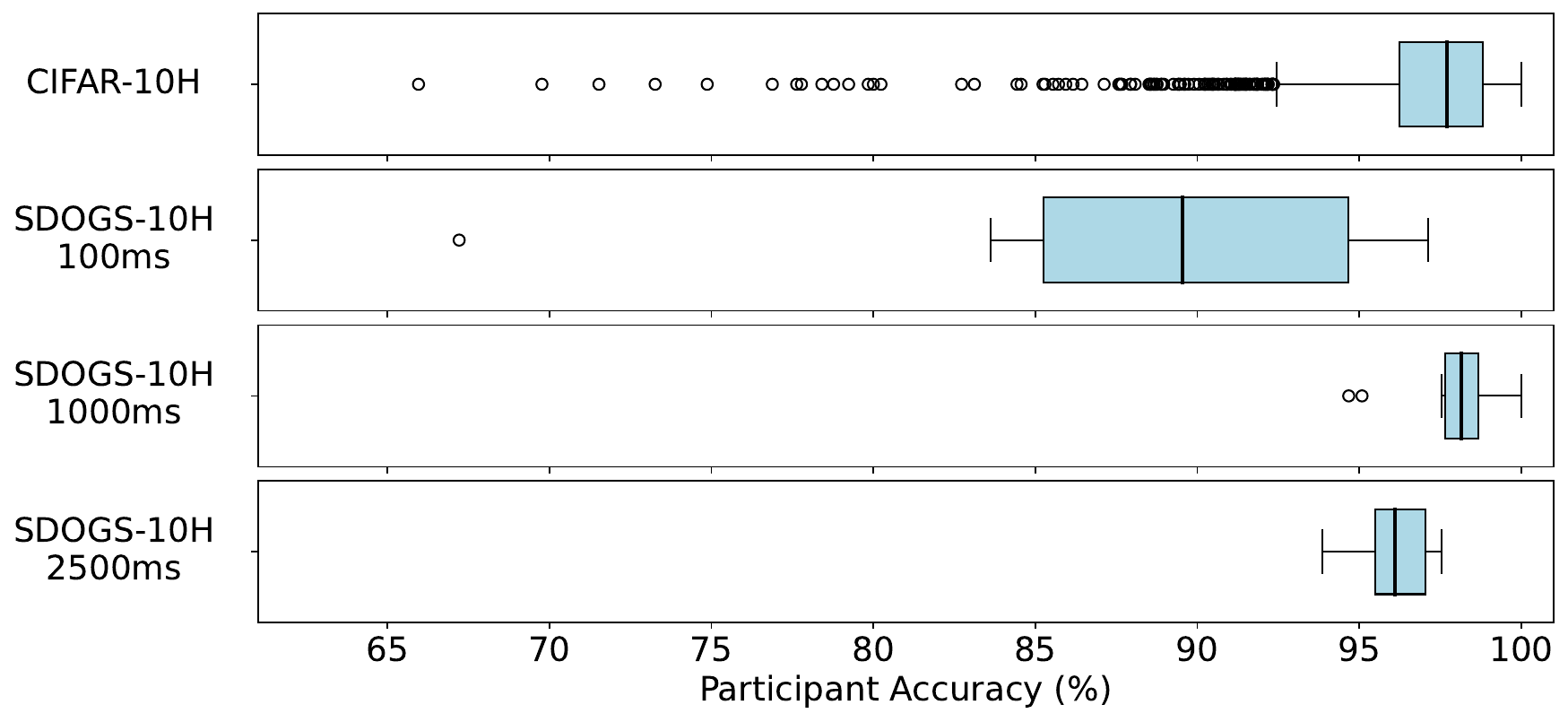}
    \caption{Participant Accuracy in CIFAR-10H and SDOGS-10H. No significant difference between CIFAR-10H and SDOGS-10H with a 1000ms view time limit suggests comparable performance at this optimal duration.}
    \label{fig:boxplot-accuracies}
\end{figure}

Figure~\ref{fig:boxplot-accuracies} shows the distribution of SDOGS-10H accuracy scores across different view time cohorts.  
Notably, the accuracy in SDOGS-10H is lowest at the 100ms view time, improves greatly at 1000ms, and surprisingly, slightly decreases again at 2500ms. 
This counterintuitive result hints at factors other than mere exposure duration such as boredom that influences participant performance in longer durations.
We then compared SDOGS-10H accuracy with those from CIFAR-10H, which uses broader and less visually similar categories such as \textit{frog}, \textit{truck}, and \textit{airplane}.
To account for the increased difficulty in SDOGS-10H, we combined the \textit{Yorkshire Terrier} and \textit{Cairn} labels, as our survey suggested that these were especially challenging for participants to differentiate (see Table~\ref{tab:code-book}).
With this adjustment, we observed that SDOGS-10H generally yields lower accuracy distributions than CIFAR-10H at the 100ms and 2500ms view times, but not at the 1000ms view time.
To statistically evaluate these differences, we conducted a two-tailed Mann-Whitney U test, comparing the accuracy distributions for each view time (100ms, 1000ms, 2500ms) in SDOGS-10H with those of CIFAR-10H.
Given that three comparisons were made, we used Bonferroni-adjusted p-values.
Table~\ref{tab:mannwhitney_results} summarizes the results. 
The p-values for the 100ms and 2500ms groups are below 0.05.
This, along with the lower means than CIFAR-10H, allows us to conclude that participants in the 100ms and 2500ms view times performed worse than participants in CIFAR-10H.
On the other hand, for the 1000ms group, we have \( p = 0.973 > 0.05\), indicating no significant difference.
Therefore, we conclude that in the 1000ms group, participants have comparable performance accuracy to those in CIFAR-10H, which did not have a time limit.
Taken together, these results suggest \emph{there is a tradeoff when introducing a time limit}, and that \emph{there exists a time limit at which performance is comparable among participants with and without a time limit imposed}.

We acknowledge that the datasets being compared differ in their categories, and that participants in the CIFAR-10H study were recruited from Amazon Mechanical Turk, which could have influenced the participant pool's quality~\cite{douglas2023data, Abbas_Gadiraju_2022}. 
However, it is important to note that our work is not intended as a direct benchmark against the previous study. 
Instead, we aim to provide insights into the relative performance of crowd workers under time-limited conditions compared to those without such constraints.

\begin{table}[t]
    \centering
    \small
    \caption{Mann-Whitney U Test Results for CIFAR-10H Participant Accuracy vs SDOGS-10H Participant Accuracy for Different View Time Limits}
    \begin{tabular}{@{}lcccc@{}}
        \toprule
        & \textbf{Cohort} & \textbf{Mean Acc} & \textbf{U Statistic} & \textbf{p-value} \\ 
        \midrule
        & 100  & 88.32\% & 22740.0 & 5.00e-05 \\
        & 1000 & 97.83\% & 10426.0 & \textbf{0.978} \\
        & 2500 & 95.98\% & 18798.5 & 0.0270 \\
        \bottomrule
    \end{tabular}
    \label{tab:mannwhitney_results}
\end{table}

\begin{figure}[htbp]
    \centering
    \includegraphics[width=0.8\linewidth]{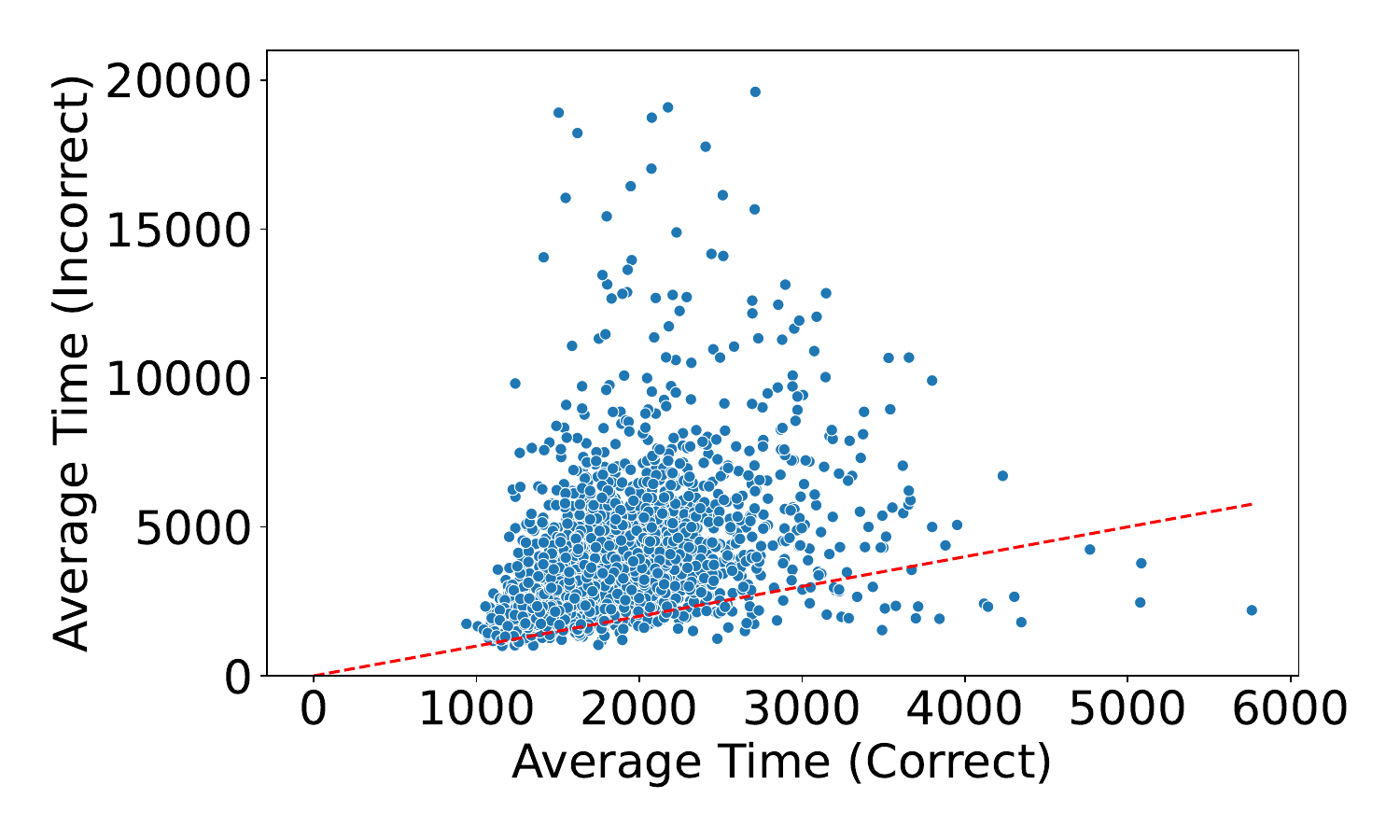}
    \caption{Average Time Taken on Incorrect vs. Correct Answers in CIFAR-10H. Crowd workers more often took longer on images they got incorrect on.
    Note: 4 points ($<0.2\%$ of data) fall outside y-axis range for clarity.}
    \label{fig:correct-vs-incorrect}
\end{figure}

\paragraph{RQ2: What is the trade-off between overall performance accuracy and varying time limits?}

Individual differences in crowd workers may lead to some requiring more time than others.  
Thus, we investigate the impact of imposing time limits on overall crowd worker performance. 
Although CIFAR-10H did not impose a time limit, they did collect time taken by participants to provide an answer.
Thus, we analyze this data to explore the effects on overall crowd worker performance. 

We first examine the relationship between time spent answering and the likelihood of providing a correct answer.
Figure~\ref{fig:correct-vs-incorrect} shows a scatter plot depicting the relationship between the average time participants spent on their correct and incorrect answers, where each point represents a participant.
The majority of data points lying above the diagonal  (2425 above and 118 below) led us to conclude that participants more often took longer on images they got incorrect on.
This observation hints at potential optimization opportunities. 
The presence of challenging images suggests that, despite investing more time, some images remain prone to label error. 
Consequently, in such instances, there are minimal opportunity costs associated with restricting the time allotted to crowd workers.
Informally, there may be cases that crowd workers are bound to identify incorrectly, regardless of time spent, so imposing a time limit will not affect their overall accuracy throughout the task. 

\begin{figure}[t]
    \centering
    \includegraphics[width=0.9\linewidth]{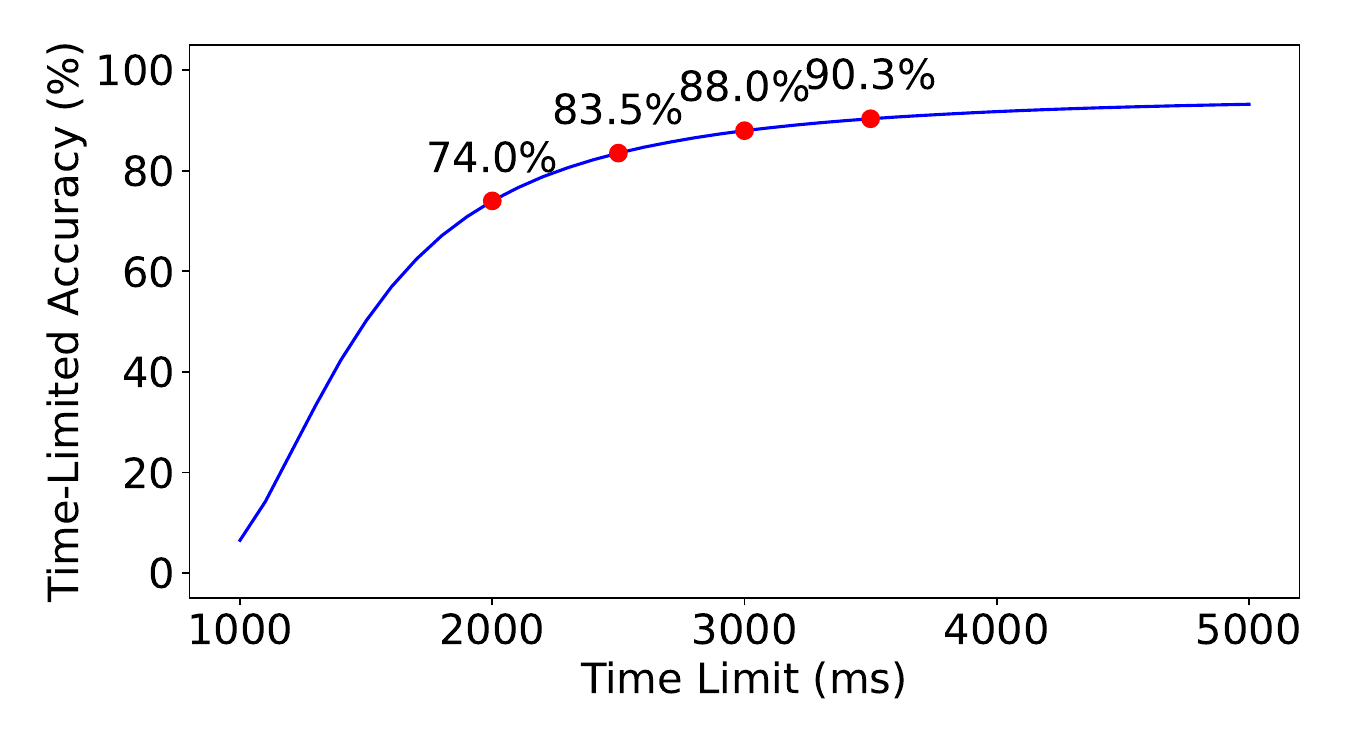}
    \caption{CIFAR-10H accuracy with answers beyond time limits (1000 to 5000ms in 100ms intervals) marked incorrect. Only 4 points labeled to avoid congestion. The graph indicates diminishing performance improvements over time.}
    \label{fig:time-limited-accuracy}
\end{figure}

Figure~\ref{fig:time-limited-accuracy} presents the overall accuracy of participants in the CIFAR-10H study while considering answers beyond hypothetical time limits as incorrect. 
The graph's shape suggests that as the duration of the time limit increases, the improvements in overall performance accuracy exhibit diminishing returns.
Informally, this assumes that individuals who spent more time on a task would have provided an incorrect answer for that task if they had been required to submit within a time limit --- this represents a worst-case scenario in terms of performance and label quality. 
Thus, we anticipate an even more pronounced diminishing returns effect with our proposed view time limit in practice.
Moreover, the ``knee'' shape in the graph suggests there is a reasonable time limit that balances the participant accuracy (and resulting data quality) against the time spent responding to tasks overall. 
\textit{Setting a view time limit leads to a diminishing impact on overall performance accuracy with increasing view time}.

\begin{figure*}[t]
    \centering
    \includegraphics[width=0.7\linewidth]{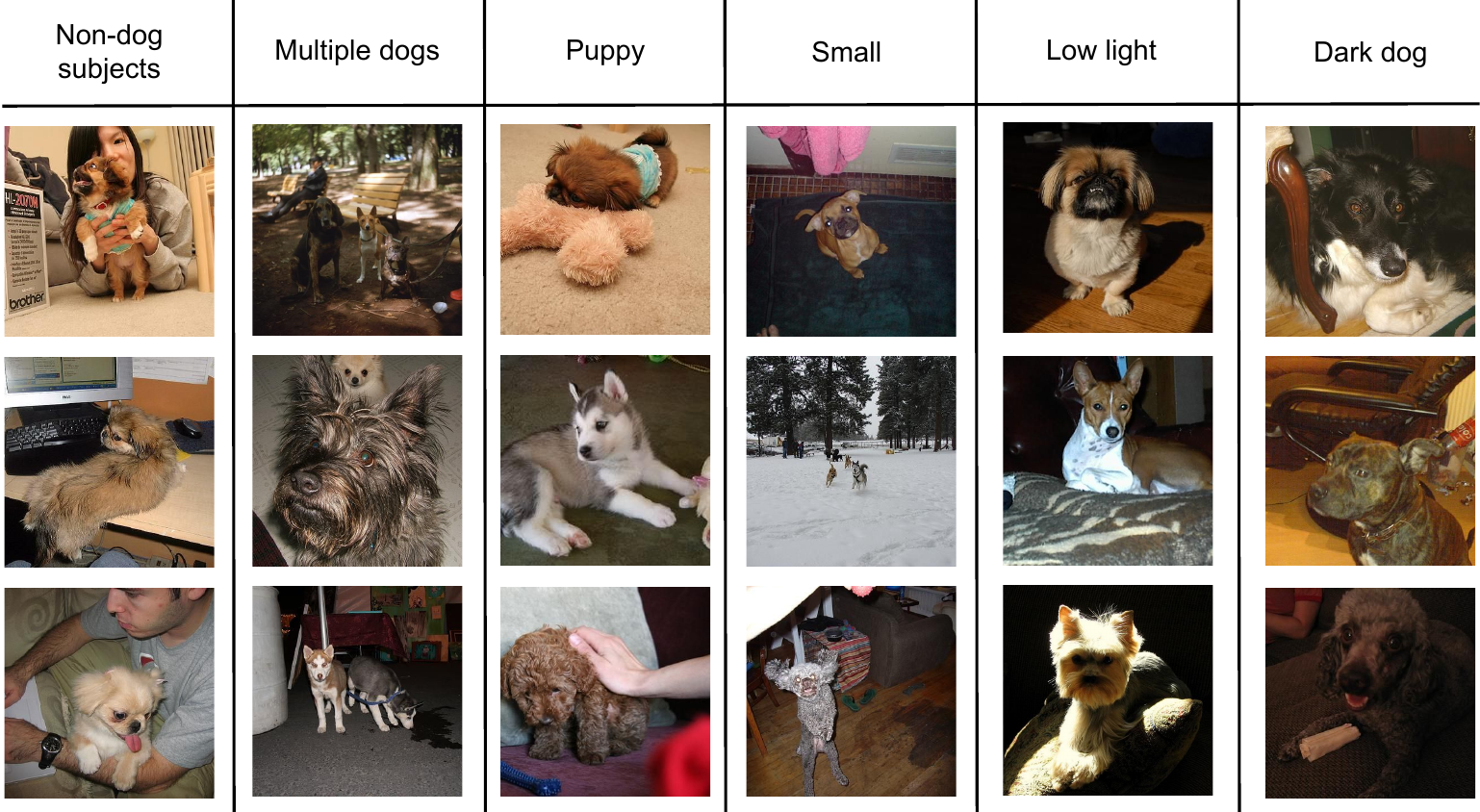}
    \caption{Select images categorized by characteristics identified as challenging under a view time limit by participants in SDOGS-10H. Note: images may exhibit multiple challenging characteristics.}
    \label{fig:difficult-images}
\end{figure*}

\begin{figure}[htbp]
    \centering
    \includegraphics[width=\linewidth]{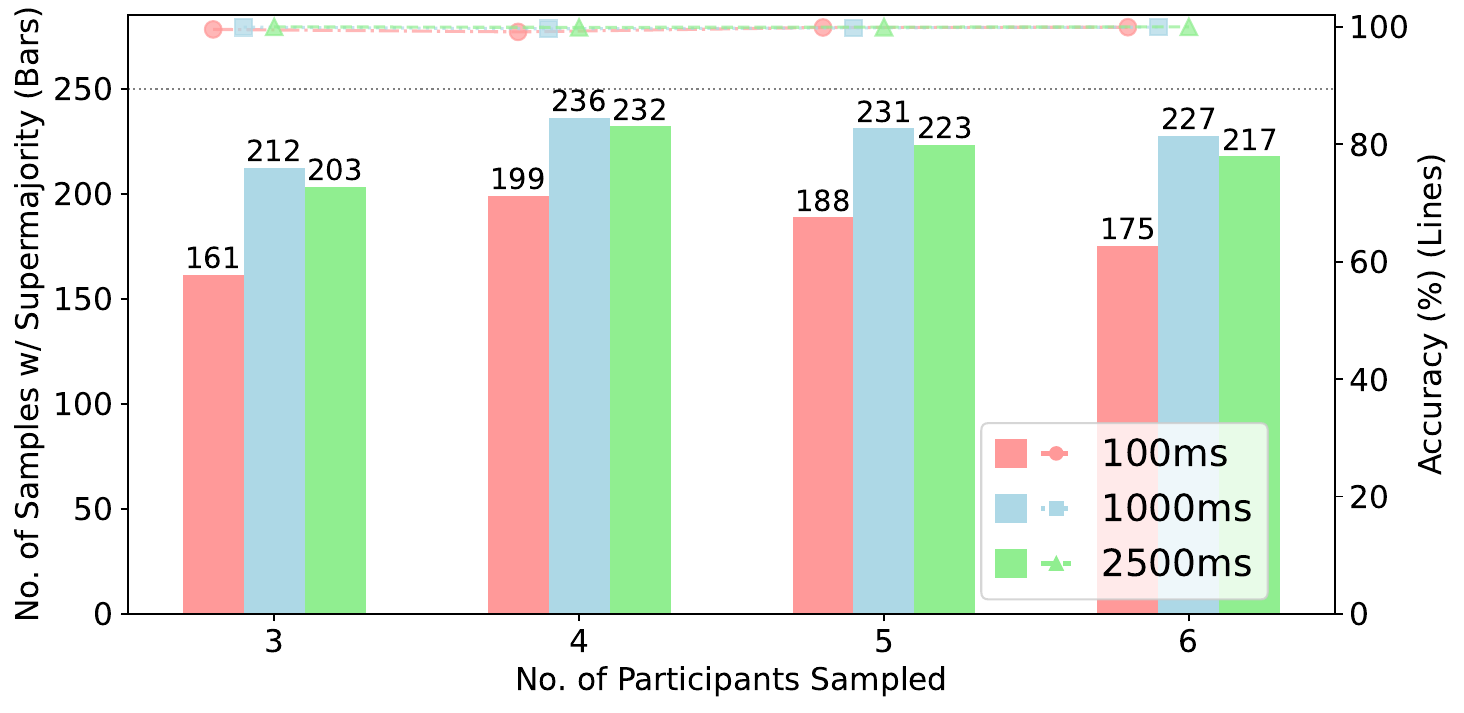}
    \caption{Number of samples with two-thirds majority (supermajority) consensus and accuracy of consensus labels (Table \ref{table:majority-accuracy}). Accuracy scores presented in Table \ref{table:majority-accuracy} for clarity.}
    \label{fig:majority-accuracy}
\end{figure}

\paragraph{RQ3: Which images are more difficult under a view time limit?}

By identifying the types of images that are especially challenging under a view time limit, we can anticipate potential failure modes within this system.
In our survey, participants were asked which types of images (i.e., from the Stanford Dogs images) were particularly difficult to classify under a time limit.
We manually curated their responses and identified several challenging characteristics, including:
subjects other than dogs, 
multiple dogs, 
puppies, 
dogs appearing small in the frame, 
low-light conditions, and
dark-colored dogs.

\begin{table}[t]
    \centering
    \small
    \caption{Average Participant Accuracy on SDOGS-10H images with problematic characteristics. 
    Images with subjects too small or far away are most susceptible to noisy labels under short view time limits.}
    \begin{tabular}{lcccc}
    \toprule
    \textbf{Chars. (Num)} & \textbf{100} & \textbf{1000} & \textbf{2500} & \textbf{Overall} \\
    \midrule
    Mult-subj (29) & 80.0\% & 99.0\% & 96.2\% & 91.7\% \\
    Mult-dog (10) & 80.0\% & 97.0\% & 97.0\% & 91.3\% \\
    Puppy (33) & 82.7\% & 95.2\% & 94.2\% & 90.7\% \\
    Small (16) & 68.8\% & 95.0\% & 88.1\% & 84.0\% \\
    Low light (21) & 86.2\% & 98.6\% & 96.2\% & 93.7\% \\
    Dark dog (44) & 88.6\% & 95.7\% & 93.4\% & 92.6\% \\
    \bottomrule
    \end{tabular}
    \label{tab:difficult-accuracy}
\end{table}

We next consider whether challenging image characteristics highlighted by participants corresponded with their performance --- that is, if participants identified an image characteristic as a source of difficulty, did they tend to make mistakes on the relevant images? 
To investigate this, we manually reviewed our test images and annotated them based on the difficult characteristics identified in the survey. 
Subsequently, we calculated the average accuracy of participants on all images sharing a specific characteristic. 
We present a selection of images with each characteristic in Figure~\ref{fig:difficult-images} and their corresponding average accuracy in Table~\ref{tab:difficult-accuracy}.

Across all characteristics, the accuracy is generally similar in the 2500ms and 1000ms cohorts, but there is a noticeable decrease in the 100ms cohort. 
For instance, the accuracy for images with multiple subjects dropped 19.0\% from 99.0\% at 1000ms and 16.2\% from 96.2\% at 2500ms to 80.0\% at 100ms, respectively. 
For images with multiple dogs, the accuracy decreased 17.0\% from 97.0\% at both 2500ms and 1000ms to 80.0\% at 100ms. 
However, for images with subjects appearing small in the frame, the decrease was notably greater, dropping 26.2\% from 95.0\% at 1000ms and 19.3\% from 88.1\% at 2500ms to 68.8\% at 100ms, respectively. 
Based on these observations, \emph{qualitative feedback can help guide the identification of challenging samples in a time-limited human intelligence task.}

\paragraph{RQ4: How can consensus algorithms mitigate the impact of time limits on performance accuracy?}

In a crowdsourced labeling task, there is an inherent risk of label noise. 
To mitigate this, it is common practice to have multiple crowd workers label each image and use a majority vote to determine the final label~\cite{northcutt2021labelerrors}.
As discussed in RQ3, a set time limit might be insufficient for some image types (e.g., small subjects).
Prior research suggests that humans may make ``fast and frugal" cognitive heuristics~\cite{homo-heuristicus}, raising concerns that even a majority might allow noisy labels to bypass a consensus algorithm if such heuristics lead people to make the same incorrect answer under a time limit.
Therefore, we investigate the effectivenes of a consensus algorithm at mitigating the impact of a time limit on crowd workers' performance.
We sampled 3 to 6 participants and kept only samples that received a two-thirds majority (or supermajority) vote for the same breed label, then calculated the accuracy of the final labels. 
We repeated this ten times and present the averaged results in Figure~\ref{fig:majority-accuracy} and Table~\ref{table:majority-accuracy}.
These data suggest that enforcing a supermajority consensus, even with just three participants, yields an improvement in overall accuracy compared to the mean individual accuracy across all three cohorts.
Additionally, for the same number of participants sampled, the number of samples with a supermajority increases from the 100ms group (161 for 3 participants) to the 1000ms group (212 for 3 participants), with only a slight drop from the 1000ms group to the 2500ms group (203 for 3 participants). 
In light of these observations, we note that
\textit{a consensus algorithm can (1) effectively preserve data quality and mitigate the impacts of a time limit and task difficulty on individual performance, and (2) filter out samples that may require another run with a higher time limit or the recruitment of more participants.}

\begin{table}[t]
\centering
\small
\caption{Accuracy of two-thirds majority consensus labels.}
\begin{tabular}{|c|r|r|r|r|}
\hline
\multirow{2}{*}{\textbf{View time (ms)}} & \multicolumn{4}{c|}{\textbf{\# Sampled Participants}} \\ \cline{2-5} 
                                        & \multicolumn{1}{c|}{\textbf{3}} & \multicolumn{1}{c|}{\textbf{4}} & \multicolumn{1}{c|}{\textbf{5}} & \multicolumn{1}{c|}{\textbf{6}} \\ \hline
100                                     & 99.57\%      & 99.15\%      & 99.89\%      & 99.94\%      \\ \hline
1000                                    & 99.95\%      & 99.66\%      & 99.83\%      & 99.96\%      \\ \hline
2500                                    & 100.00\%      & 99.91\%      & 99.96\%      & 100.00\%      \\ \hline
\end{tabular}
\label{table:majority-accuracy}
\end{table}

\paragraph{RQ5: How does a view time limit affect the effort and satisfaction of crowd workers over the duration of our task?}

\begin{figure}[h]
    \centering
    \includegraphics[width=\linewidth]{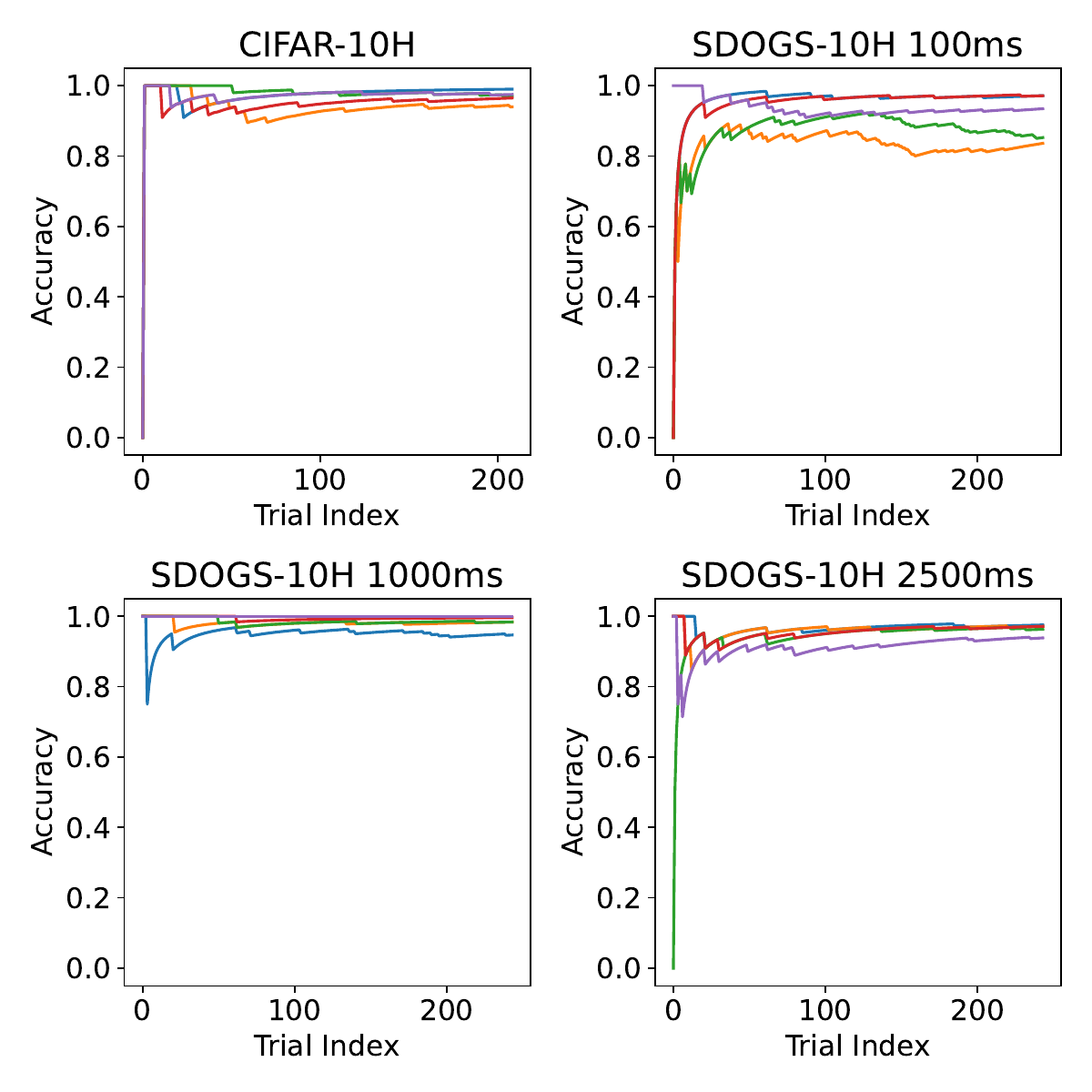}
    \caption{Participants' accuracy over their trial. Each line represents a sampled participant. 
    Compared to CIFAR-10H participants without a time limit, SDOGS-10H participants with a view time limit maintained good accuracy.}
    \label{fig:accuracy-over-trial}
\end{figure}

A time-limited task must remain favorable and manageable for crowd workers to work on such that they are not deterred from choosing to work on the task or expect higher compensation, while still maintaining consistent effort throughout.
We examine the accuracy of participants over the duration of the test. 
For each cohort in our study as well as the CIFAR-10H study, we analyzed five sampled participants and present the results in Figure~\ref{fig:accuracy-over-trial}.
In the CIFAR-10H setup, where no time limit was enforced, accuracy remained consistently high, with minor early fluctuations. 
This pattern persisted across the three view time limit cohorts in our study.
However, we note that two participants in the 100ms cohort showed a decrease in accuracy around the same point. 
Nevertheless, we conclude that the imposed view time limits did not substantially affect participant performance as the test progressed, and they maintained consistent effort throughout. 

Referring again to the survey responses (Table~\ref{tab:code-book}), participants who perceive the duration as too short and, as a result, challenging might be less satisfied with their work, feeling that they could have performed better with more time. 
On the other hand, a few subjects in our longest view time cohort (2500ms) indicated that the enforced view time limit was too long, and expressed that they would have preferred to have been able to submit their responses earlier. 
This suggests that they may view the task less favorably due to the imposed view time limit.
Nevertheless, across all view time limit cohorts, there are participants who have left positive comments for our study.

\begin{figure}[h]
    \centering
    \includegraphics[width=0.7\linewidth]{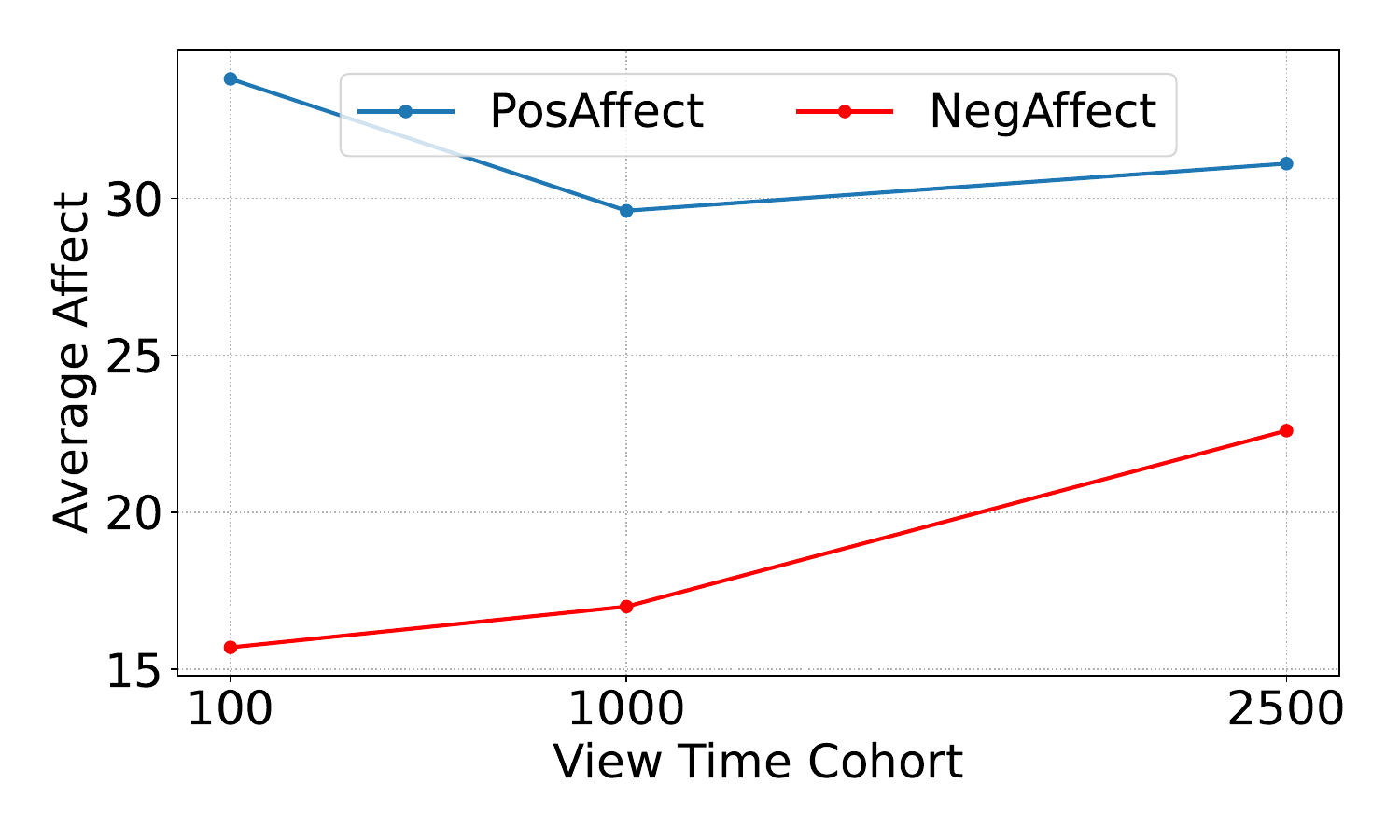}
    \caption{Average Positive affect (PosAffect) and Negative affect (NegAffect) scores of participants in the different view time cohorts. Decreasing differences between average PosAffect and NegAffect scores suggests crowd workers perceive shorter view time limits more preferably.}
    \label{fig:affect-scores}
\end{figure}

Figure~\ref{fig:affect-scores} presents the average PANAS scores of participants in the different view time cohorts.
The difference between the Positive Affect (PosAffect) scores and Negative Affect (NegAffect) scores decreases with increasingly long view time limits, suggesting that participants prefer a shorter time limit. 
We conducted a Kruskal-Wallis test to compare the differences in affect scores among the three viewtime groups (100ms, 1000ms, and 2500ms). 
The test indicated a statistically significant difference among the groups ($H = 736.96$, $df = 2$, $p < 0.001$).
To identify which specific groups differed from each other, we performed Dunn's post hoc test.
this revealed a statistically significant difference between Positive and Negative Affect Scores among all three pairs of groups (100ms vs. 1000ms, 100ms vs. 2500ms, and 1000ms vs. 2500ms), all with $p < 0.001$.
Combined with the increasing average difference in affect scores with increasingly long view time limits, we are led to conclude that when a time limit is imposed, workers prefer a shorter time limit.
A plausible reason is that with a shorter view time limit, participants undertake a lighter cognitive load. 
Additionally, participants may perceive a degree of tolerance for uncertainty within the study parameters, thereby experiencing reduced stress levels.

\section{Discussion and Conclusion}

Our analysis marks the first comprehensive examination of the impact of a view time limit in image classification crowdsourcing tasks within real-world data collection pipelines.
We have shown that while time limits can indeed negatively impact accuracy, this impact increasingly diminishes with longer view times. 
Additionally, while some types of images are especially challenging under a time limit, a consensus algorithm remains effective at preserving data quality and filtering samples that need longer times.
We also showed there were no major fatigue effects in our study as evidenced by their consistent performance throughout the task. 
Finally, when a time limit is imposed, workers prefer shorter time limits, as suggested by their PANAS scores.

Subsequently, we detail some recommendations for the implementation of a view time limit beyond our experimental setting:
\begin{itemize}
    \item Allow crowd workers to submit their answers before their view time limit expires to prevent dissatisfaction and performance degradation caused by delays as supported by~\citet{time-delay}.
    \item Conduct preliminary lab trials to determine an appropriate view time limit duration, which will be essential in managing the impact to individual crowd worker performance~\cite{measuring-crwdsrc-effort}.
    \item Use consensus scoring, such as a majority vote~\cite{northcutt2021labelerrors}, among multiple crowd workers to mitigate the impact to individual crowd worker performance variation and preserve data quality.
\end{itemize}

A limitation of our study is its restriction to tasks that prompt crowd workers on a target image.
Future research could explore analogous mechanisms in more complex tasks, such as object segmentation, to establish reasonable time limits and assess their feasibility.
One challenge is limiting cognitive processing time while allowing sufficient time for the psychomotor task of physically submitting the answer~\cite{measuring-crwdsrc-effort}, which we performed by limiting only the image view time and not the time to submit an answer.
However, for tasks like object segmentation, it is not feasible to make the image they are annotating disappear.
Additionally, the psychomotor costs of annotation for object segmentation may have a greater impact on completion times.

In summary, our work highlights the viability of imposing time limits in crowdsourced tasks to better predict task timing and to better manage participant expectations. 
Our data shows how there is a balance between time limit duration and participant accuracy, but that there is a \emph{sweet spot} that can be used to co-optimize participant performance against total time taken.
By adopting this approach, we encourage responsible crowdsourcing methodologies that better ensure fair and transparent payment.

\begin{acks}
ChatGPT was used to assist this Work by revising portions of text and generating code snippets, which were subsequently reviewed and refined to ensure accuracy and quality.
\end{acks}

\section*{Ethical Considerations}
The implementation of a time limit for image classification, as we have proposed, along with our concluding recommendations, aims to promote ethical practices, fair compensation, and transparency in crowdsourcing. 
However, we recognize that if misused, this approach could inadvertently lead to unethical consequences.
First, while our proposed time limit allows the crowd worker to submit their response and receive compensation even after the limit has elapsed, it is important that the task itself does not "time out" without payment, thereby disregarding the effort the worker has already invested~\cite{toxtli2021quantifying}.
Second, the time limits we have explored for image classification tasks are relatively short, typically only a few seconds, and they only begin when the crowd worker initiates each image task. 
This design does not noticeably interfere with the worker's ability to take breaks or ``task switch" between images~\cite{lascau2022crowdworkers, crowdsrc-on-call-2023}. 
However, if these time limits were extended to tasks such as processing long documents, which may require several minutes, the worker could be compelled to focus on a single task for the entire duration, potentially limiting their temporal flexibility.

\bibliographystyle{ACM-Reference-Format}
\bibliography{wsdm/wsdm25}

\appendix

\end{document}